\documentclass[aip,reprint,amsmath,amssymb]{revtex4-2}

\usepackage{fontenc}
\usepackage{anyfontsize}
\usepackage{dblfloatfix}
\usepackage{indentfirst}
\usepackage{calc}
\usepackage{xcolor}
\usepackage{graphicx}
\usepackage{subcaption}
\usepackage{dcolumn}
\usepackage{bm}
\usepackage{ifthen}
\usepackage{soul}
\usepackage{upgreek}
\usepackage{svg}
\usepackage{soul}
\usepackage{xr-hyper}
\usepackage[hidelinks,colorlinks]{hyperref}

\DeclareUnicodeCharacter{2212}{$-$}
\definecolor{ref}{HTML}{3DB4E5}
\DeclareMathOperator\sinc{sinc}
\newcommand\matr[1]{\bm{#1}}

\renewcommand\deg{^\circ}
\newcommand\refsupSecProtocol{Appendix A }
\newcommand\refsupSecSystematics{Appendix C }
\newcommand{\supplement}[1]{\href{https://doi.org/10.1063/5.0166136}{#1}}

\hypersetup{
    allcolors=black,
    citecolor=ref,
    linkcolor=ref,
    urlcolor=ref
}

\begin{document}

\preprint{APS/123-QED}

\title{Mass and shape determination of optically levitated nanoparticles}
\author{Bart Schellenberg}
\author{Mina Morshed Behbahani}
\author{Nithesh Balasubramanian}
\author{Ties H. Fikkers}
\author{Steven Hoekstra}\email{s.hoekstra@rug.nl}
\affiliation{Van Swinderen Institute, University of Groningen, The Netherlands and Nikhef, Amsterdam, The Netherlands}
\date{\today}

\begin{abstract}
When introducing a nanoparticle into an optical trap, its mass and shape are not immediately apparent. We combine a charge-based mass measurement with a shape determination method based on light scattering and an analysis of the damping rate anisotropy, all on the same set of silica nanoparticles, trapped using optical tweezers in vacuum. These methods have previously only been used
separately, and the mass determination method has not been applied to asymmetric particles before. We demonstrate
that the combination of these classification techniques is required to distinguish particles with similar mass but
different shape, and vice versa. The ability to identify these parameters is a key step for a range of experiments on precision measurements and sensing using optically levitated nanoparticles.
\end{abstract}

\maketitle

\begin{figure}[th]
    \centering
    \begin{subfigure}{.25\textwidth}
        \vspace*{-.5cm}\hspace*{-.8cm}\def\svgscale{0.1545}
\begingroup%
  \makeatletter%
  \providecommand\color[2][]{%
    \errmessage{(Inkscape) Color is used for the text in Inkscape, but the package 'color.sty' is not loaded}%
    \renewcommand\color[2][]{}%
  }%
  \providecommand\transparent[1]{%
    \errmessage{(Inkscape) Transparency is used (non-zero) for the text in Inkscape, but the package 'transparent.sty' is not loaded}%
    \renewcommand\transparent[1]{}%
  }%
  \providecommand\rotatebox[2]{#2}%
  \newcommand*\fsize{\dimexpr\f@size pt\relax}%
  \newcommand*\lineheight[1]{\fontsize{\fsize}{#1\fsize}\selectfont}%
  \ifx\svgwidth\undefined%
    \setlength{\unitlength}{885bp}%
    \ifx\svgscale\undefined%
      \relax%
    \else%
      \setlength{\unitlength}{\unitlength * \real{\svgscale}}%
    \fi%
  \else%
    \setlength{\unitlength}{\svgwidth}%
  \fi%
  \global\let\svgwidth\undefined%
  \global\let\svgscale\undefined%
  \makeatother%
  \begin{picture}(1,0.74576271)%
    \lineheight{1}%
    \setlength\tabcolsep{0pt}%
    \put(0,0){\includegraphics[width=\unitlength,page=1]{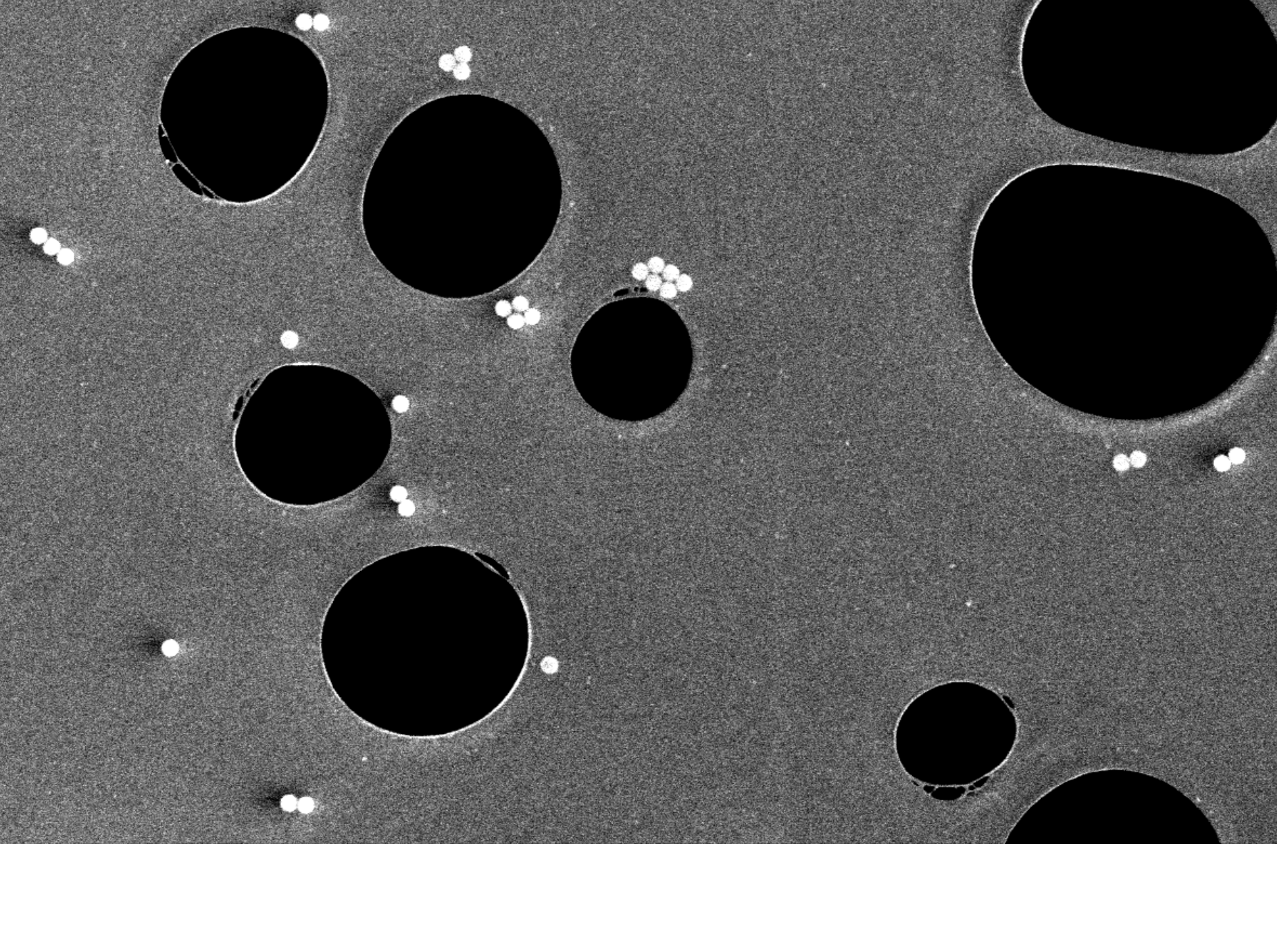}}%
    \put(0.90192803,0.02421251){\color[rgb]{0,0.82745098,0.82745098}\makebox(0,0)[t]{\lineheight{1.25}\smash{\begin{tabular}[t]{c}1 µm\end{tabular}}}}%
    \put(0,0){\includegraphics[width=\unitlength,page=2]{semzoo.pdf}}%
    \put(0.23117983,0.42930148){\color[rgb]{1,0.91764706,0}\makebox(0,0)[lt]{\lineheight{1.25}\smash{\begin{tabular}[t]{l}1\end{tabular}}}}%
    \put(0,0){\includegraphics[width=\unitlength,page=3]{semzoo.pdf}}%
    \put(0.32150293,0.50074081){\color[rgb]{1,0.91764706,0}\makebox(0,0)[lt]{\lineheight{1.25}\smash{\begin{tabular}[t]{l}4\end{tabular}}}}%
    \put(0,0){\includegraphics[width=\unitlength,page=4]{semzoo.pdf}}%
    \put(0.27388106,0.695818){\color[rgb]{1,0.91764706,0}\makebox(0,0)[lt]{\lineheight{1.25}\smash{\begin{tabular}[t]{l}3\end{tabular}}}}%
    \put(0,0){\includegraphics[width=\unitlength,page=5]{semzoo.pdf}}%
    \put(0.14736468,0.11774145){\color[rgb]{1,0.91764706,0}\makebox(0,0)[lt]{\lineheight{1.25}\smash{\begin{tabular}[t]{l}2\end{tabular}}}}%
  \end{picture}%
\endgroup%
        \vspace*{-.082cm}
    \end{subfigure}
    \begin{subfigure}{.15\textwidth}
        \vspace*{-.5cm}\hspace*{-.2cm}\def\svgscale{.8}
\begingroup%
  \makeatletter%
  \providecommand\color[2][]{%
    \errmessage{(Inkscape) Color is used for the text in Inkscape, but the package 'color.sty' is not loaded}%
    \renewcommand\color[2][]{}%
  }%
  \providecommand\transparent[1]{%
    \errmessage{(Inkscape) Transparency is used (non-zero) for the text in Inkscape, but the package 'transparent.sty' is not loaded}%
    \renewcommand\transparent[1]{}%
  }%
  \providecommand\rotatebox[2]{#2}%
  \newcommand*\fsize{\dimexpr\f@size pt\relax}%
  \newcommand*\lineheight[1]{\fontsize{\fsize}{#1\fsize}\selectfont}%
  \ifx\svgwidth\undefined%
    \setlength{\unitlength}{120.40199661bp}%
    \ifx\svgscale\undefined%
      \relax%
    \else%
      \setlength{\unitlength}{\unitlength * \real{\svgscale}}%
    \fi%
  \else%
    \setlength{\unitlength}{\svgwidth}%
  \fi%
  \global\let\svgwidth\undefined%
  \global\let\svgscale\undefined%
  \makeatother%
  \begin{picture}(1,1.06229133)%
    \lineheight{1}%
    \setlength\tabcolsep{0pt}%
    \put(0,0){\includegraphics[width=\unitlength,page=1]{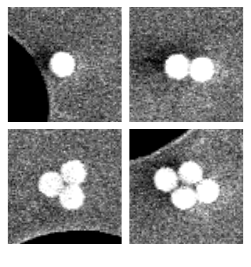}}%
    \put(0.04160016,0.95542437){\color[rgb]{1,0.91764706,0}\makebox(0,0)[lt]{\lineheight{1.25}\smash{\begin{tabular}[t]{l}\textbf{1}\end{tabular}}}}%
    \put(0.52602789,0.47099727){\color[rgb]{1,0.91764706,0}\makebox(0,0)[lt]{\lineheight{1.25}\smash{\begin{tabular}[t]{l}\textbf{4}\end{tabular}}}}%
    \put(0.04160015,0.47099727){\color[rgb]{1,0.91764706,0}\makebox(0,0)[lt]{\lineheight{1.25}\smash{\begin{tabular}[t]{l}\textbf{3}\end{tabular}}}}%
    \put(0.52602596,0.9554248){\color[rgb]{1,0.91764706,0}\makebox(0,0)[lt]{\lineheight{1.25}\smash{\begin{tabular}[t]{l}\textbf{2}\end{tabular}}}}%
    \put(0.80712786,0.00341129){\color[rgb]{0,0.82745098,0.82745098}\makebox(0,0)[t]{\lineheight{1.25}\smash{\begin{tabular}[t]{c}200 nm\end{tabular}}}}%
    \put(0,0){\includegraphics[width=\unitlength,page=2]{semfour.pdf}}%
  \end{picture}%
\endgroup%

    \end{subfigure}
    \caption{Two pictures revealing possible shapes of the (composite) nanoparticles used for this paper, taken using a scanning electron microscope. The black patches represent holes in the holey carbon substrate. The left picture shows some of the nanoparticles used for trapping, where each sphere has a diameter of $142\pm4\,\text{nm}$, according to the manufacturer. The panels on the right show a close-up of several composite structures.\label{fig:particles}}
\end{figure}

With a rapidly increasing number of developments over the recent years, levitated nanospheres have evolved into an exciting platform for innovative measurement opportunities and applications. Demonstrated applications span from the manipulation of microscopic biological systems\cite{10.1021/cr4003006,10.1140/epje/i2005-10060-4,10.1016/s0006-3495(97)78780-0,10.1007/s12551-019-00599-y} to ultra-sensitive accelerometers and force-sensors,\cite{Gieseler.Quidant.2013,Hempston.Ulbricht.2017,Lewandowski.D’Urso.2021,Ranjit.Geraci.2016,10.1103/physrevlett.121.063602} torque detectors,\cite{10.1103/physrevlett.121.033603,10.1038/s41565-019-0605-9,10.1103/physrevlett.92.190801,10.1080/23746149.2020.1838322} hyper-fast mechanical rotors,\cite{10.1038/ncomms3374,Reimann.Novotny.2018zhdp,Jin.Zhang.2021} and measurements on thermal diffusion.\cite{Li.Raizen.2010,Bellando.Louyer.2022,Gieseler.Millen.2018,Hoang.Li.2018,Li.2013} Numerous proposals in recent years have explored the potential of using isolated nanometer-sized particles for probing gravitational waves,\cite{Arvanitaki.Geraci.2013,Marshman.Bose.2020} to observe quantum gravity,\cite{Margalit.Folman.2021,10.1103/physreva.102.062807,10.1103/physrevlett.119.240402} to employ in dark matter scattering experiments,\cite{10.1103/physrevlett.128.101301,10.1103/physrevlett.125.181102} or to trial quantum collapse models.\cite{10.1103/physreva.100.012119} For numerous of these\cite{10.1103/physrevlett.121.063602, 10.1103/physrevlett.121.033603,10.1038/s41565-019-0605-9,10.1103/physrevlett.92.190801,10.1080/23746149.2020.1838322, 10.1103/physrevlett.128.101301,10.1103/physrevlett.125.181102,Ranjit.Geraci.2016} and other\cite{Gonzalez-Ballestero.Romero-Isart.2021,Stickler.Kim.2021h7,10.1002/adma.200901271,10.1007/s40820-015-0040-x,10.1016/j.jare.2010.02.002} applications, knowing the precise mass and morphology of the levitated particles is essential. When introducing a particle into an optical tweezer however, its shape and mass are not immediately apparent. Particles from a monodisperse solution of spheres have been observed to regularly carry some non-negligible ellipticity,\cite{Rademacher.Barker.2022} or they may aggregate to form composite structures.\cite{10.1103/physrevlett.121.033603}\\
\indent While mass determination of charged particles and shape determination through the anisotropy of light scattering and the damping rate have been demonstrated on an individual basis, a comprehensive correlative study of these techniques applied to a set of differently shaped and sized particles remains unexplored. In this work, we present the shape and mass determination for optically levitated silica nanoparticles of various sizes and shapes in vacuum. We demonstrate that the combination of these classification techniques is required to discriminate between different particles. We manage to capture both single particles as well as aggregated structures in an optical tweezer trap by adjusting the concentration of a monodisperse solution of nanospheres.\cite{10.1103/physrevlett.121.033603} As illustration of some of the typical shapes that we encounter, \autoref{fig:particles} shows a scanning electron microscope picture of our solution. To unambiguously determine the mass and shape of the optically levitated nanoparticles we combine the above-mentioned in situ classification methods.\\
\indent Specifically, we extend the mass determination that was previously demonstrated for nanospheres\cite{10.1021/acs.nanolett.9b00082,ricci2019levitodynamics} to asymmetric compositions of nanospheres, using similar and smaller sizes. We combine this method with the shape determination from a non-isotropic damping rate due to residual background gas.\cite{10.1103/physrevlett.121.033603,ThesisAhn2020} In addition, by employing a secondary probe laser, we find the particle's morphology through its angle-resolved Rayleigh scattering profile.\cite{Rademacher.Barker.2022} We discuss the reliability of each method individually as well as their combined results. Our approach does not rely on precise calculations of the moment of inertia or the polarisability of the particle.\\
\;\\
\indent The structure of this paper is as follows. Following an introduction of the experimental setup, we present the methods and results from each individual classification technique to determine the properties of a fixed set of particles. After that we correlate and discuss the combined results.\\
\begin{figure}[ht]
    \centering
    \def\svgwidth{8.5cm}
    {\normalfont\fontsize{2.8mm}{1}\selectfont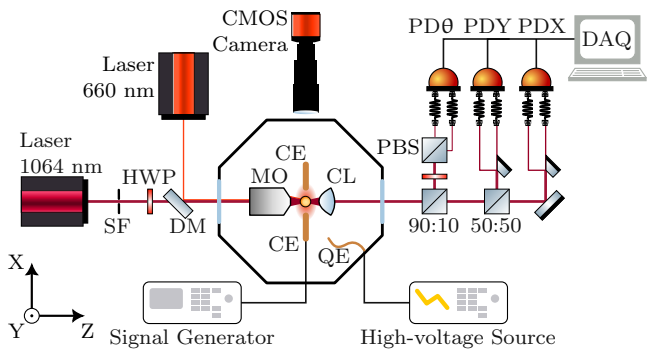}
    \caption{A schematic representation of the experimental setup. Details are presented in the main text.\label{fig:setup}}
\end{figure}
\;\\
\;\\
\;\\
\indent We combine a number of standard techniques\cite{Gieseler.Novotny.2012vwa} to optically trap nanoparticles in a controlled pressure environment. A schematic of our experimental setup is shown in \autoref{fig:setup}. We use the light of a Coherent Nd:YAG Mephisto MOPA trapping laser ($\lambda_{\text{trap}}=1064\,\text{nm}$) and a spatial filter (SF) to obtain a Gaussian trapping beam. Using a half-waveplate (HWP) we control its polarization. The light then enters the vacuum chamber through a window and is focused using a microscope objective (MO; $\text{NA}=0.8$), establishing the dipole trap at its focus point. The strong and inhomogeneous electric field at the focus creates a gradient force on the dielectric nanoparticles, allowing them to be trapped at the focus. We evapourate silica ($\text{SiO}_2$) nanospheres (diameters $103\pm6,\,142\pm4\,\text{nm}$; Microparticles GmbH) from an ethanol solution using a medical nebulizer\cite{10.1364/oe.16.007739} at ambient pressure near the trapping region. When a droplet from the cloud of ethanol, carrying a nanoparticle, passes through the optical trap, the particle can be caught. Once the particle is trapped, we reduce the pressure toward the operational domain between $\sim20$ and $\sim0.1\,\text{mbar}$.\\
\indent Surrounding the trapping region, we have placed two copper electrodes (CE), with which we create a controlled electric field at the location of the trapped particle. We placed a discharge electrode (QE) to be able to charge/discharge the trapped nanoparticle.\cite{10.1021/acs.nanolett.9b00082,ricci2019levitodynamics} The electrode setup is used in the mass measurement of the nanoparticle. To visualize the trapped nanoparticle, and to measure its angular Rayleigh scattering profile, we use the light of a probe diode laser ($\lambda_{\text{probe}}=660\,\text{nm}$), and overlap this beam with the trapping beam using a dichroic mirror (DM). A fraction of the scattered light at $\lambda_{\text{probe}}$ is collected using a CMOS camera. To track the dynamics of the particle inside the trap, we collect the transmitted trapping light using a collection lens (CL), and guide it toward a series of beamsplitters. We first use $10\%$ of the remaining light for the angular detection, by employing a polarizing beam splitter (PBS) and a differential photodiode (PD$\uptheta$), which is balanced by another HWP. The final $90\%$ is used for the X and Y detection, which include a D-shaped mirror and a differential photodiode (PDX and PDY), to measure the spatial intensity differences within the beam profile.\\
\indent For each particle whose data are represented in this paper, the same experimental procedure was conducted. A detailed description on this procedure is given in \refsupSecProtocol of the \supplement{supplementary material}. The particle is essentially cleaned from possible residual water in the porous internal silica structure,\cite{10.1016/s0927-7757(00)00556-2,ricci2019levitodynamics,10.1039/c5sm02772a} and charged to about $6$-$10$ charges in preparation of the mass measurement. Once ready, we perform the mass measurements and record the angular Rayleigh scattering profile in the harmonic regime at around $20\,\text{mbar}$. We then periodically record the particle's signal during a slow pump-down, to determine its damping rate as a function of pressure.\\
\begin{figure}[ht]
    \centering
    \vspace*{-.5cm}\hspace*{.3cm}\def\svgwidth{8.5cm}
    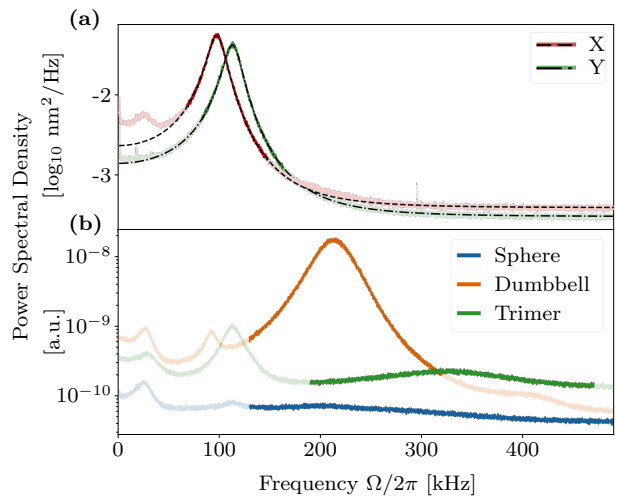
    \caption{Typical power spectral densities (PSD) of the trapped nanoparticles. \textbf{(a)} The transverse (X,Y) motion of a spherical nanoparticle (particle 15 from \autoref{fig:result}). The dashed/dotted lines indicate the best-fit of \autoref{eq:PSD}; \textbf{(b)} PSDs of the signal recorded by PD$\uptheta$ for three particles (15, 18, and 6 from \autoref{fig:result}); The additional peaks in both insets correspond to the Z, X, and Y motions of the particle. (See \refsupSecSystematics of the \supplement{supplementary text} for further details.) The highlighted parts in both insets indicate the regions where the signal of interest dominates, which are used for the analysis.\label{fig:PSD}}
    {\phantomsubcaption{\label{fig:PSDa}}}
    {\phantomsubcaption{\label{fig:PSDb}}}
\end{figure}
\;\\
\indent When the translational damping of the particle is sufficiently high ($\gtrsim3\,\text{mbar}$), its motion is primarily described by the (single-sided) linear power spectral density (PSD)\cite{ThesisGieseler2014,ThesisErick2017}
\begin{align}
    S_F(\Omega)&=\frac{k_BT}{m}\frac{\Gamma_0}{(\Omega_0^2-\Omega^2)^2+\Omega^2\Gamma_0^2}\label{eq:PSD},
\end{align}
which can be directly obtained from the signals recorded by PDX and PDY. Here $\Omega_0$ represents the natural oscillation frequency and $\Gamma_0$ is the damping rate of the nanoparticle due to the background gas, both in $\text{rad}/\text{s}$. The mass of the particle is denoted with $m$, and $k_B$ and $T$ are the Boltzmann constant and the heatbath temperature respectively. The subscript $F$ is used to denote that \autoref{eq:PSD} is driven by the thermal fluctuation force. \autoref{fig:PSDa} shows a typical translational PSD from a spherical nanoparticle. The cyclic frequencies of the X and Y channels are not degenerate due to some ellipticity in the optical trap.\cite{Jin.Zhang.2019} We obtain the natural frequency $\Omega_0$ and the damping rates $\Gamma_0$ by fitting \autoref{eq:PSD} to \autoref{fig:PSDa}. To obtain an impression of the ellipticity of the particle's morphology, we also use the torsional PSD, recorded by PD$\uptheta$. Typical torsional PSDs for three different particles are shown in \autoref{fig:PSDb}. The three particles were classified as a nanosphere, a dumbbell, and a triangle trimer, using the techniques described in this paper. We observe similar trap frequencies $\Omega_0$ (and therefore similar trap stiffness) for the translational motion of these three particles.\\
\begin{figure}[b]
    \centering
    \vspace*{-.5cm}\hspace*{.15cm}\def\svgscale{0.4}
    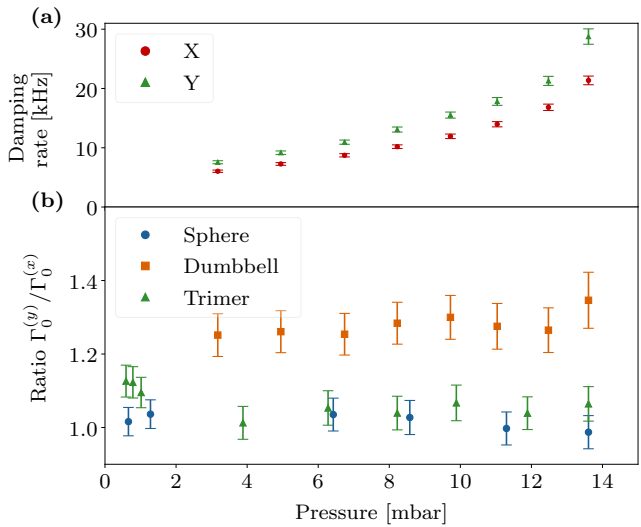
    \caption{\textbf{(a)} The damping rate of a nanoparticle as a function of pressure, corresponding to particle 3 from \autoref{fig:result}; \textbf{(b)} the ratio $\Gamma_0^{(y)}/\Gamma_0^{(x)}$ for a set of different particles, which were classified as a sphere, a dumbbell, and a triangle trimer. The particles correspond to 15, 3, and 6 from \autoref{fig:result} respectively.\label{fig:damping}}
    {\phantomsubcaption{\label{fig:dampingRate}}}
    {\phantomsubcaption{\label{fig:dampingRatio}}}
\end{figure}
\indent In linearly polarized light, the asymmetric susceptibility tensor of a non-spherical particle will introduce a torque to the system that causes the particle to align its major axis of polarizability with the polarization of the trapping light. One of the consequences of this laser-induced alignment is that the measured damping rates in the X and Y channels become unequal. This anisotropy in the damping rates can be used as a measure of the asymmetry of the particle's shape.\cite{10.1103/physrevlett.121.033603,ThesisAhn2020} \autoref{fig:dampingRate} shows the damping rates of a nanoparticle, which was classified as a dumbbell, as a function of the pressure. \autoref{fig:dampingRatio} shows the corresponding ratio $\Gamma_0^{(y)}/\Gamma_0^{(x)}$, as well as that of a sphere and a triangle trimer. The nanodumbbell shows a high degree of asymmetry, with a ratio $\Gamma_0^{(y)}/\Gamma_0^{(x)}\approx1.27$, which is in agreement with that of a dumbbell with a length to diameter ratio close to 1.7.\cite{10.1103/physrevlett.121.033603} Meanwhile, the sphere ($\Gamma_0^{(y)}/\Gamma_0^{(x)}\approx1.03$) and trimer ($\Gamma_0^{(y)}/\Gamma_0^{(x)}\approx1.11$) appear to be much more symmetric in the XY plane of the setup, as is in agreement with simulated results,\cite{ThesisAhn2020} and they require additional classification techniques to be unambiguously distinguished from one another.\\
\indent In addition to the asymmetric damping rates, we also record the angle-resolved Rayleigh scattering profile of the particle, which was recently demonstrated to allow the detection of asymmetries in the particle's morphology down to a few nanometers\cite{Rademacher.Barker.2022} in the XY-plane of our setup. To realize this procedure, we use the probe laser (co-propagating with the trapping laser along the Z-axis) to illuminate the nanoparticle inside the optical trap at a different wavelength ($\lambda_{\text{probe}}=660\,\text{nm}$) and lower power ($\sim2\,\text{mW}$) from the trapping laser ($\lambda_{\text{trap}}=1064\,\text{nm}$; $\sim330\,\text{mW}$). A HWP is used to rotate the linear polarization of the trapping laser, while keeping that of the probe laser fixed. The laser induced optical alignment of an asymmetric particle then allows us to effectively rotate it about the longitudinal Z axis of our setup. The scattered light is then detected at a right angle using a $12$-bit CMOS camera, from which we track the averaged intensity over a closed pixel region around the particle. To improve the signal-to-noise ratio, we image the nanoparticle slightly out of focus, such that we could adjust the gain of the camera without causing the pixels to immediately saturate. A dichroic mirror is used to prevent the scattered light at the trapping wavelength from reaching the camera.\\
\indent Theoretically, the expected scattering intensity can be obtained from the particle's susceptibility tensor as follows:\cite{Rademacher.Barker.2022}
\begin{align}
    I(\theta)\propto\big(\chi^{(yy)}\big)^2;\;\;\;\text{with}\;\;\;\matr{\chi}(\theta)=\matr{R}\matr{\chi}_0\matr{R}^\intercal\label{eq:scat},
\end{align}
where $\matr{\chi}_0$ represents the susceptibility tensor in the particle's eigenframe, and $\matr{R}$ is the rotation matrix to map to the laboratory frame. For the case of a dumbbell specifically, values for $\chi_0^{(yy)}/\chi_0^{(xx)}$ have been computed,\cite{10.1103/physrevlett.121.033603} and suggest $I_{\text{min}}/I_{\text{max}}\approx0.75$ when the length to diameter ratio reaches close to 1.7. \autoref{fig:scattering} shows the resulting data for three particles, which were classified as a nanosphere, a dumbbell, and a triangle trimer. The susceptibility tensor of the particle in \autoref{fig:scattering_sphere} appears to be almost fully symmetrical in the XY-plane, however, the data reveal some small ellipticity of the particle at $I_{\text{min}}/I_{\text{max}}\approx0.99$. In contrast, the particle in \autoref{fig:scattering_dumbbell} shows a high degree of asymmetry between $\chi_0^{(yy)}$ and $\chi_0^{(xx)}$, with $I_{\text{min}}/I_{\text{max}}\approx0.72$, and most likely resulted from a dumbbell. The particle in \autoref{fig:scattering_trimer}, however, shows far less contrast and could not be distinguished from a single elliptical nanosphere based on the scattering data alone. By including its mass and damping rate measurements, this particle was classified as a triangle trimer.\\
\begin{figure}[ht]
    \centering
    \vspace*{-.5cm}\hspace*{.5cm}\def\svgwidth{8.5cm}
    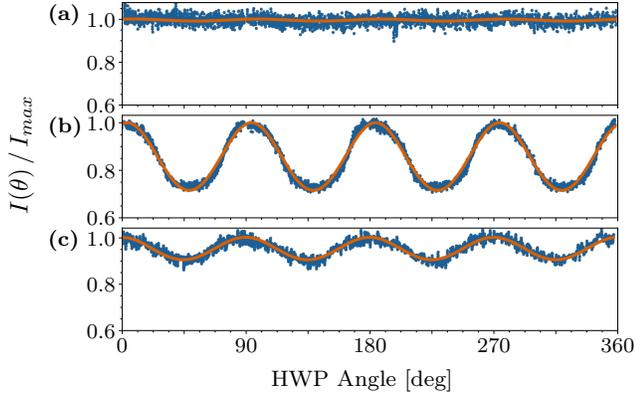
    \caption{Measured Rayleigh scattering profile for several different particles. Each particle shown in this figure consisted of a number of nanospheres, with an individual size of $d=103\pm6\,\text{nm}$. The data were taken from particles 1, 3, and 6 from \autoref{fig:result} respectively.}\label{fig:scattering}
    {\phantomsubcaption{}\label{fig:scattering_sphere}}
    {\phantomsubcaption{}\label{fig:scattering_dumbbell}}
    {\phantomsubcaption{}\label{fig:scattering_trimer}}
\end{figure}
\;\\
\;\\
\indent The instantaneous inertial response of the nanoparticle to a well defined externally applied force may be used to directly determine the mass of the particle. A protocol for the active control over the nanoparticle's electric charge has already been developed,\cite{ricci2019levitodynamics} and is also elaborated on in \refsupSecProtocol of the \supplement{supplementary material}. We use the CE to harmonically drive the charged particle using a quasi-static electric field near its natural oscillation frequency. Due to the zero correlation between the electrical driving force and the random Brownian fluctuation force, which is responsible for \autoref{eq:PSD}, the PSD of this driven system can be written by simply appending the additional term\cite{10.1021/acs.nanolett.9b00082}
\begin{align}
    S_M(\Omega)&=\frac{F_0^2\tau}{8m^2}\frac{\sinc\big(\frac12(\Omega-\Omega_M)\tau\big)}{(\Omega_0^2-\Omega^2)^2+\Omega^2\Gamma_0^2}\label{eq:PSDMass},
\end{align}
to \autoref{eq:PSD}. In this new term, $F_0$ represents the amplitude of the applied sinusoidal force, $\Omega_M$ the frequency of this force, and $\tau$ is the duration of the finitely-recorded signal used to compute the Fourier transform. The subscript $M$ is used to denote that \autoref{eq:PSDMass} results from the modulation force that is applied to the particle. While \autoref{eq:PSD} scales with the inverse of the particle's mass $m$, \autoref{eq:PSDMass} does with its square. Therefore, the ratio
\begin{align}
    \frac{S_F(\Omega_M)}{S_M(\Omega_M)}=\frac{8\Gamma_0k_BT}{F_0^2\tau}m,
\end{align}
may be used to extract the particle's mass.\\
\indent \autoref{fig:PSDMassSphere} and \autoref{fig:PSDMassDumbbell} show two examples of the PSD following two harmonically driven nanoparticles, which were classified as a sphere and a dumbbell. The inset in both figures shows a close-up of the PSD around the driving frequency $\Omega_M$, revealing the sinc-shape as described by \autoref{eq:PSDMass}. The dashed lines on both show the corresponding fits to \autoref{eq:PSD} for the large-scale figure and \autoref{eq:PSDMass} for the inset. Both particles are driven using the exact same force, and the presented data were recorded at the same pressure ($\sim15\,\text{mbar}$). Nonetheless, it can be seen that \autoref{fig:PSDMassDumbbell} has a lower peak and a smaller overall width than \autoref{fig:PSDMassSphere}, the latter of which relates to a lower damping rate $\Gamma_0$. Both properties indicate that \autoref{fig:PSDMassDumbbell} was taken from a heavier particle than \autoref{fig:PSDMassSphere}. Below, in \autoref{fig:PSDMassResult}, the resulting mass is shown for both particles in femtograms. To improve our statistics, we measure each particle for a series of driving frequencies $\Omega_M$ around the particle's natural frequency $\Omega_0$. From the results, it can be seen that the mass of the dumbbell is about twice as much as that of a single sphere.\\
\begin{figure}[h]
    \centering
    \vspace*{-.5cm}\def\svgscale{.4}
    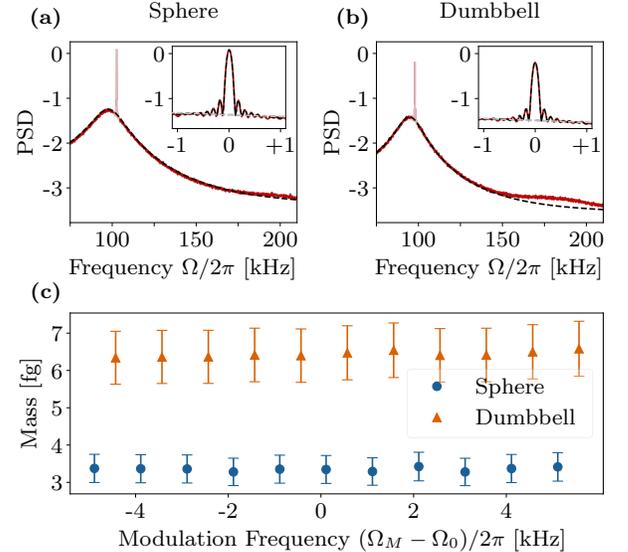
    \caption{Mass data for two particles. \textbf{(a)} and \textbf{(b)} show the PSDs following two harmonically driven particles, alongside an enlarged version of the driving peak in the inset, following \autoref{eq:PSDMass}. The dashed lines in both figures represent the best-fit to \autoref{eq:PSD} and \autoref{eq:PSDMass}. In \textbf{(b)}, one may observe that part of the torsional signal is leaked into the data starting around $180\,\text{kHz}$. \textbf{(a)} and \textbf{(b)} correspond to particles 15 and 17 in \autoref{fig:result}. \textbf{(c)} shows the resulting mass measurements for both particles, measured for a series of driving frequencies $\Omega_M$ around the natural frequency $\Omega_0$.\label{fig:PSDMass}}
    {\phantomsubcaption{\label{fig:PSDMassSphere}}}
    {\phantomsubcaption{\label{fig:PSDMassDumbbell}}}
    {\phantomsubcaption{\label{fig:PSDMassResult}}}
\end{figure}
\;\\
\indent We will now consider the combined results of the damping rate, scattering, and mass measurements. \autoref{fig:result} shows an overview of a set of 18 different nanoparticles. For each particle, we have performed the three classification methods, which leads us to categorize as indicated at the bottom of the figure. The horizontal dashed lines represent the average results and the standard deviation of the spread, for each classification category.\\
\indent As a single classification technique, the mass measurements appear to show the best resolution. We classify particles whose masses are approximately twice as much as that of the corresponding spheres as dumbbells, and those which are three times as much as trimers. We find that the mass of a nanoparticle is typically a bit high compared to the specifications of the manufacturer. One likely reason for this is the presence of residual water in the silica structure.\cite{10.1016/s0927-7757(00)00556-2,ricci2019levitodynamics,10.1039/c5sm02772a} However, also uncertainties in the mass density of the nanoparticles play a role.\cite{ricci2019levitodynamics,10.1039/c5sm02772a} (See also \refsupSecProtocol in the \supplement{supplementary material} for more details.) Our results on the mass determination are in line with those obtained in other publications.\cite{10.1021/acs.nanolett.9b00082,ricci2019levitodynamics}\\
\begin{figure}[ht]
    \centering
    \vspace*{-.5cm}\def\svgscale{.4}
    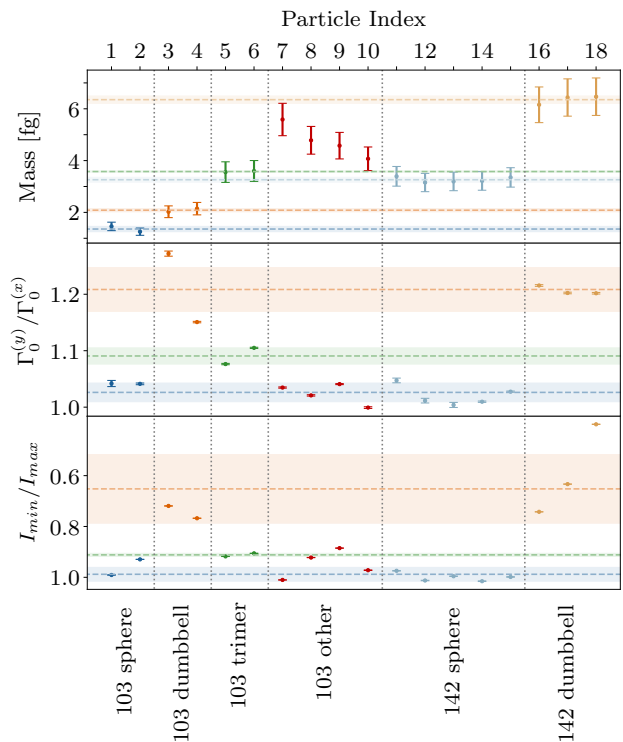
    \caption{The combined results from the mass, damping rate, and scattering measurements, sorted by their classified categories. The data on each column corresponds to the same particles. The horizontal dashed lines represent the average result per category (excluding "other", which represents compositions of four or more spheres), and the colored band shows the standard deviation of the data points.}\label{fig:result}
\end{figure}
\;\\
\indent The measurements on the scattering profile and the damping rates do not show a significant difference between the nanoparticles composed of $103\,\text{nm}$ or $142\,\text{nm}$ spheres, as these methods focus on the morphology of the particle. Neither does either of the methods single-handedly give a satisfactory resolution that allows for an unambiguous distinction between single spheres, triangle trimers, and particles composed of four or more nanospheres (denoted as "other" in \autoref{fig:result}). In this case, the combination of multiple classification techniques is required to classify the particle.\\
\indent On a single particle level, the measurements on the mass and shape using the three methods (PSD analysis, scattering anisotropy and mass determination) are reproducible and show consistent results. When comparing multiple particles of the same category with each other, however, we see significant scattering of the data beyond the statistical error bars, which lead us to perform an analysis of the main systematic effects that influence these measurements. Here, we present the main conclusions, while the full details are given in the \supplement{supplementary material}.\\
\indent Concerning the mass measurement, the accuracy with which we can determine the distance between the copper electrodes limits the accuracy with which we can quantify the mass of the particle. In our results this is given as a symmetric uncertainty of $5.6\%$ in the value of the electric field. The fitting of the PSD, which is used to assess the morphology of the particle, is influenced by the crosstalk between the different motional modes of the particle. At pressure below a few mbar, where damping by background gas is reduced, the anharmonic part of the potential is explored which leads to a reduced quality of the fits. We find an overestimation of the damping rate of a given motional degree of freedom in the pressure range we use by at most $5\%$. Regarding the scattering anisotropy, we have evaluated detector linearity and alignment as potential systematic effects, which play a minor role. In this paper, we have primarily demonstrated the classification of spheres, dumbbells, and triangle trimers, as these appear to be some of the simplest configurations that we most often trap from our solutions. However, nanospheres can also aggregate into different shapes, such as chain trimers, or obtuse configurations. These shapes were not observed in our measurements, which we suspect is due to the stability of our trap and the rate at which these configurations form in our solution.\\
\indent We have combined a charge-based mass measurement with a shape determination method based on light-scattering and an analysis of the damping rate anisotropy, all on the same set of trapped nanoparticles. We observe a large variation in the shapes and sizes within a set of nanoparticles. This is not only caused by the variation in the shape and mass of the individual spheres, but also by the way in which they combine to form composite particles. We have demonstrated that the combination of these classification techniques enables us to obtain an unambiguous conclusion on the particle’s shape, size, and mass.\\
\\
\indent See the \supplement{supplementary material} for a detailed description of the measurement protocol, the data analysis (specifically for obtaining the damping rate $\Gamma_0$), and the consideration of systematic effects in all three classification techniques.

\begin{acknowledgments}
We acknowledge the support from Gert ten Brink, Leo Huisman and George Palasantzas. This project has received funding from NWO through NWA Startimpuls (400.17.608/4303).
\end{acknowledgments}

\section*{Author Declarations}

\subsection*{Conflict of Interest}
The authors have no conflicts to disclose.

\subsection*{Author Contributions}
\textbf{Bart Schellenberg:} Conceptualisation (equal); Data curation (equal); Formal analysis (lead); Investigation (equal); Methodology (equal); Software (equal); Validation (equal); Visualisation (equal); Writing - original draft (equal); Writing - review \& editing (equal). \textbf{Mina Morshed Behbahani:} Conceptualisation (equal); Data curation (equal); Investigation (equal); Methodology (equal); Software (equal); Validation (equal); Visualisation (equal); Writing - original draft (equal); Writing - review \& editing (equal). \textbf{Nithesh Balasubramanian:} Conceptualisation (equal); Formal analysis (supporting); Investigation (equal); Methodology (equal); Software (equal); Writing - original draft (supporting); Writing - review \& editing (equal). \textbf{Ties H. Fikkers:} Conceptualisation (equal); Investigation (equal); Methodology (equal); Writing - original draft (supporting); Writing - review \& editing (equal). \textbf{Steven Hoekstra:} Conceptualisation (equal); Funding acquisition (lead); Investigation (equal); Methodology (equal); Validation (equal); Visualisation (equal); Writing - original draft (equal); Writing - review \& editing (equal).

\section*{Data Availability Statement}
The data that support the findings of this study are available from the corresponding author upon reasonable request.

\section*{References}
\bibliography{nanospheres.bib}

\appendix
\onecolumngrid
\pagebreak
\newcommand\paprefFigResult{\autoref{fig:result} }
\newcommand\paprefEqPSD{\autoref{eq:PSD} }

\section{Pre-measurement Protocol}\label{sec:protocol}
\begin{figure}[ht]
    \centering
    \vspace*{-.5cm}\def\svgscale{0.55}
    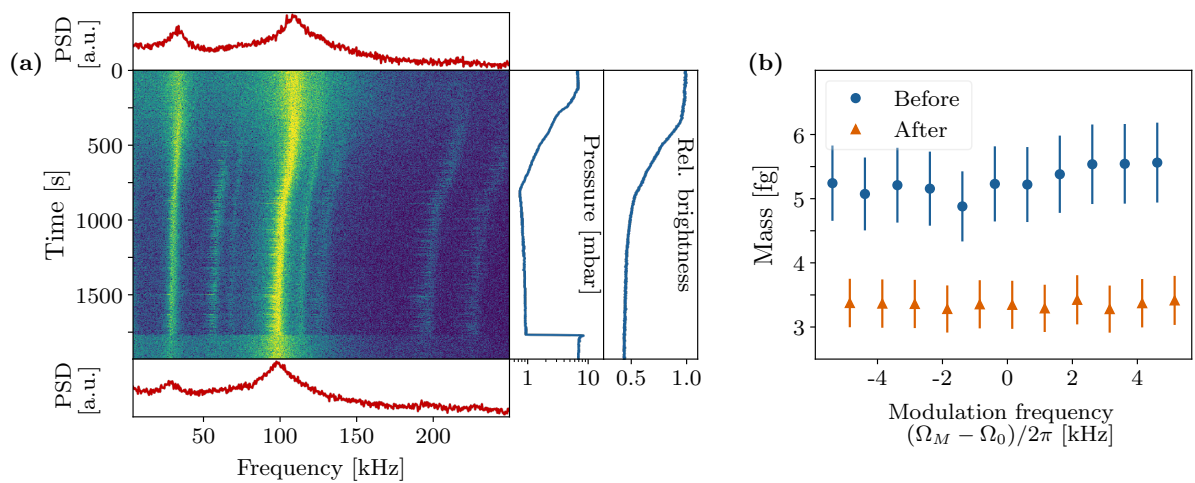
    \caption{Observed changes in the particle's composition throughout a pump-down from $\sim7\,\text{mbar}$ down to $\sim0.7\,\text{mbar}$ and back. This pump-down is a standard part of our experiment, and it enhances the consistency of our results.\label{supfig:massdrop}}
    \phantomsubcaption{\label{supfig:massdrop_waterfall}}
    \phantomsubcaption{\label{supfig:massdrop_mass}}
\end{figure}

For each particle whose data are represented in the main text, the same experimental procedure was conducted. We first mix the aqueous suspension ($5\%\,\text{w/v}$) of silica nanospheres with ethanol. By adjusting the concentrations, we alter the average number of nanospheres per ethanol droplet,\cite{ricci2019levitodynamics} allowing us to alter the probability to obtain a single particle or a composite structure. The mixture is then suspended in an ultra-sonic bath for $15\,\text{minutes}$. Once the mixture is ready, we evaporate it near the trapping region using a medical nebulizer\cite{10.1364/oe.16.007739} at ambient pressure. Occasionally, a droplet from the cloud of ethanol carries a nanoparticle toward the trapping region and leaves it there.\\
\indent Once a nanoparticle is trapped, we reduce the pressure inside the vacuum chamber to about $\lesssim1\,\text{mbar}$. Around this pressure, the transverse cyclic frequency ($\Omega_0^{(x,y)}/2\pi\approx100\,\text{kHz}$) is typically observed to drop by a few kHz. This drop appears to coincide with a drop in the total mass of the particle at the order of $\sim10\%$, but does not seem to show consistency throughout each experiment. \autoref{supfig:massdrop} shows an overview of our observations. In \autoref{supfig:massdrop_waterfall} the power spectral density (PSD) is recorded continuously over time, the top and the bottom subfigures show the initial and final PSDs. A shift of $\sim10\,\text{kHz}$ is observed during the recording. The figures on the right show the measured pressure inside the vacuum chamber, and the scattering brightness recorded by the CMOS camera.\\
\;\\
\indent \autoref{supfig:massdrop_mass} shows the mass of the same particle as it was measured before and after the pump-down, with a relative drop of about $36\%$. For demonstrative purposes, the particular particle reported here was chosen as it showed a relatively large change during the pump-down. A mass drop of $\sim12\%$ has been reported before,\cite{ricci2019levitodynamics} though the exact changes appear to be highly inconsistent between measurements. The change in the mass of the particle could be attributed to the evaporation of residual water inside the silica structure.\cite{10.1016/s0927-7757(00)00556-2,ricci2019levitodynamics,10.1039/c5sm02772a} The particle reported in \autoref{supfig:massdrop} corresponds to particle 15 in \paprefFigResult of the main paper.\\
\indent After keeping the particle at $\lesssim1\,\text{mbar}$ for some time, the vacuum chamber is vented and the pressure is brought to $\sim7\,\text{mbar}$. To couple the nanoparticle to the electric field, created using the copper electrodes (CE), we then proceed to use the charging electrode (QE) to cause a controlled discharge at around $-900\,\text{V}$ inside the vacuum chamber. This causes the nanoparticle to occasionally pick up a charge.\cite{10.1021/acs.nanolett.9b00082,ricci2019levitodynamics} To track the number of net charges that are present on the nanoparticle, we use the CE to harmonically drive the particle at some known frequency ($\Omega_M$) close to its natural oscillation frequency. The addition or removal of individual charges will then show up as discrete jumps in the Fourier spectrum at this particular frequency. The presence and precise knowledge of the (absolute) number of charges is an essential step in measuring the particle's mass.\\
\begin{figure}[ht]
    \centering
    \vspace*{-.5cm}\def\svgscale{0.5}
    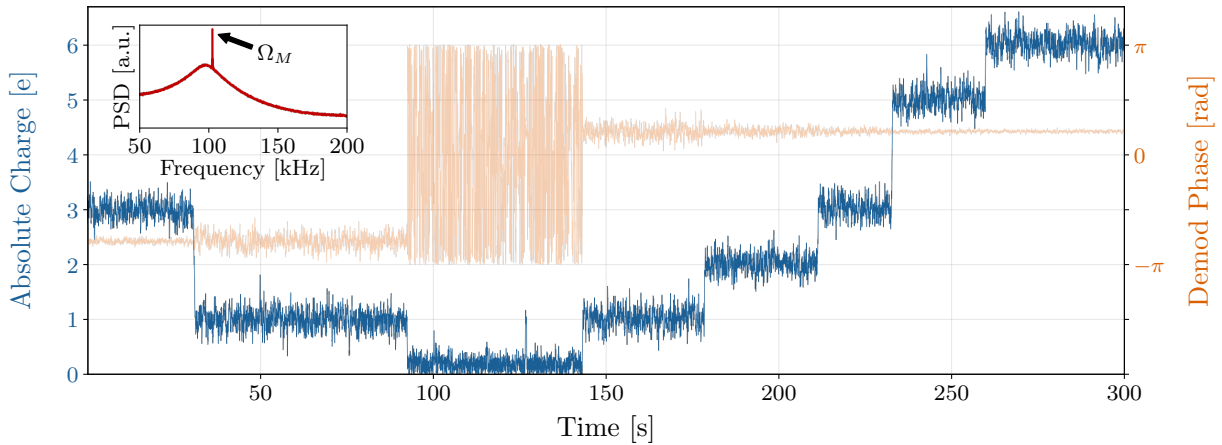
    \caption{The value of the PSD at the modulation frequency $\Omega_M$ tracked over time, normalized to show the net number of charges that are present on the nanoparticle. The 'demod phase' data in the background denotes the phase error of the demodulator in the DAQ device. In this particular example, the particle carried $3$ charges prior to the charging stage. The phase change before $90\,\text{s}$ and after $145\,\text{s}$ indicates a change in the sign of the net charge. In the intermediate interval the particle was electrically neutral, and the demodulator could not find the phase of its oscillation at $\Omega_M$, hence an error spread of $2\pi$ was recorded.\label{supfig:charging}}
\end{figure}
\;\\
\indent Once the particle carries typically around $6$-$10$ charges, the pressure is increased once again to about $20\,\text{mbar}$, at which the particle's motion is sufficiently damped to be in a fully harmonic regime. At this pressure, both the mass measurement as well as the probing of the angular scattering profile is performed. In the final stage of the experiment, the pressure is steadily brought back down again. During this pump-down, the particle's positional signal is periodically recorded, which will be used to determine the damping rate. This process continues until the lack of thermal damping due to the background gas causes the particle to heat up too much, and leave the trap. We typically loose our particles at $\sim0.1\,\text{mbar}$ with the current state of our setup.\\

\section{Data Processing}\label{sec:processing}
To obtain information about the properties of the particle, in particular its damping rate $\Gamma_0$, from the translational signal, we primarily make use of the PSD. In the harmonic regime ($\gtrsim3\,\text{mbar}$), its behaviour is primarily described by \paprefEqPSD of the main paper, which is reported here for completeness
\begin{align}
    S_F(\Omega)&=\frac{k_BT}{m}\frac{\Gamma_0}{(\Omega_0^2-\Omega^2)^2+\Omega^2\Gamma_0^2}\label{supeq:PSD}.
\end{align}
To obtain the damping rate outside the harmonic regime ($\lesssim3\,\text{mbar}$), we instead resort to the autocorrelation of the time-varying local amplitude of the particle's motion, which decreases exponentially with time following\cite{Bellando.Louyer.2022}
\begin{align}
    \langle X(t)\,X(t+\tau)\rangle_t\propto e^{-\Gamma_0\tau}\label{supeq:autocorrelation}.
\end{align}
\indent We obtain the instantaneous amplitude $X(t)$ by applying a Hilbert transform to the signal, for a finite frequency window around the particle's natural frequency. \autoref{supfig:autocorrelation} illustrates a set of autocorrelation functions obtained from the data recorded by PDX. We typically observed the autocorrelation to be more sensitive to noise and systematic effects than the PSD, an example of which can be seen toward the top-left of \autoref{supfig:autocorrelation}. Yet a satisfactory result may still be obtained, as \autoref{supeq:autocorrelation} only comes with a single degree of freedom.\\

\section{Systematic Effects}\label{sec:systematics}

\begin{figure}[ht]
    \centering
    \vspace*{-.5cm}\def\svgscale{.5}
    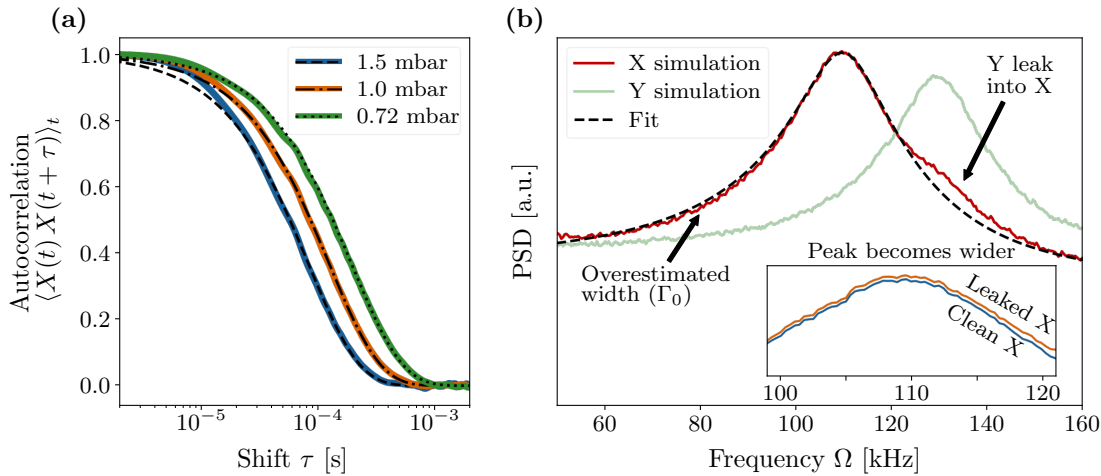
    \caption{\textnormal{\textbf{(a)}} The autocorrelation function following the instantaneous amplitude of the position data of a particle. The dashed/dotted lines represent the best fit of equation \autoref{supeq:autocorrelation}. \textnormal{\textbf{(b)}} Simulation used for determining the fitting quality. In this particular example, the width of the fit, which corresponds to the damping rate $\Gamma_0$, is overestimated due to part of the Y signal leaking in.\label{supfig:autoleak}}
    \phantomsubcaption{\label{supfig:autocorrelation}}
    \phantomsubcaption{\label{supfig:leaking}}
\end{figure}

\paragraph*{Systematics in the Damping Rate}\;\\
To obtain the best-fit parameters from both the PSD, \autoref{supeq:PSD}, and the autocorrelation, \autoref{supeq:autocorrelation}, we apply a standard least-squares fit to our data. We determined the uncertainty of our fitting procedures by employing a series of simulations.\cite{Nørrelykke.Flyvbjerg.2011} The different channels of the simulated data were mixed as to mimic any leakage of the signals due to a potential misalignment of the detection setup. Subsequently, a white-noise field was applied to mimic the thermal and electronic effects in the detection. We then tried to recover the initial conditions used to simulate the data set, while continuously varying the leaking and noise levels.\\
\indent From the simulations we concluded that the damping rate was typically overestimated by between $0\%$ and $5\%$ due to misalignments, as the signal of one channel was partially leaked into the detection of another, causing its energy to be appended on top of the PSD. This misalignment could be attributed to the imperfect placement of the photodiodes relative to each other, or relative to the ellipticity of the beam profile.\cite{Jin.Zhang.2019} \autoref{supfig:leaking} shows an example of such a situation, where part of the Y channel of a simulated set is leaked into the X detection. Not only does the bump at $\Omega_0^{(y)}$ cause the fitting algorithm to overestimate the width of the distribution, the distribution itself is also widened, as can be seen from the inset figure.\\
\;\\
\paragraph*{Systematics in the Particle Formation}\;\\
We are not entirely sure exactly when and how the composite particles are formed. Usually we observe the trapping of a nano-dumbbell to be instantaneous, typically indicated by a greater brightness on the CMOS camera than a single sphere. However, occasionally after trapping a particle with a lower brightness, we have also observed a sudden jump in the brightness, suggesting that a secondary particle may have entered the trap. These composite particles do not seem to behave in any different way as compared to particles that were immediately trapped as a whole, and whose structure might have already formed at an earlier stage. The high and non-linear sensitivity of the optical properties of a composite particle on the individual separation-distance of its spheres\cite{10.1103/physrevlett.121.033603} might account for a relatively large spread in the damping and scattering rate results for the nano-dumbbell and triangle trimer, as seen in \paprefFigResult of the main paper.\\
\;\\
\paragraph*{Systematics in the Scattering Profile}\;\\
To determine the angle-resolved Rayleigh scattering profile of the trapped nanoparticles, we employed a secondary probe laser ($\lambda_{\text{probe}}=660\,\text{nm}$) to illuminate the particle, and we used a CMOS camera to record its scattering intensity at a $90\deg$ angle w.r.t. the optical (Z) axis. The light of the probe laser was combined with the trapping beam ($\lambda_{\text{trap}}=1064\,\text{nm}$) using a longpass dichroic mirror (DM; $900\,\text{nm}$ cut-on), which was optimized to transmit the trapping laser and reflect the probe laser. At $\lambda_{\text{trap}}$ the DM still has a reflection coefficient of about $1.7\%$. We used this residual $1.7\%$ of the trapping light, in combination with a polarizing beam splitter and a photodiode, to track the polarization angle of the trapping laser after the half-waveplate (HWP).\\
\indent Our systematic effects in measuring the scattering profile of the nanoparticle primarily show up in the angle-reconstruction of the polarization of the trapping laser, which tends to show difficulties especially at integer multiples of $\theta_{\text{HWP}}=45\deg$, where the angular-derivative of the intensity measured by the photodiode reaches zero. Additionally, a low scattering brightness occasionally poses an issue when trying to distinguish the scattered light from the background noise, causing our signal-to-noise ratio to decrease. We did not observe any changes in the scattering profile of the same nanoparticle when slightly adjusting the optical alignment, by for instance changing the overlap of the probe laser onto the trapping beam, or when slightly moving the trap itself by carefully moving the trapping laser around.\\

\paragraph*{Systematics in the Mass Measurements}\;\\
The driving force that is applied to the nanoparticle by the CE directly relates to the electric field at the trapping region following $F_0=n_q\,e\,E_0$, where $n_q$ is the number of electric charges, $e$ the elementary charge unit, and $E_0$ the electric field. To determine $E_0$, a series of finite element method simulations was performed using COMSOL. We found that the primary source of uncertainty in the measured mass relates to the separation distance of the CE. This almost solely determines the total uncertainty in our mass measurements, and it represents a systematic shift (quadratic with the particle's mass) that is the same for each particle in this paper.\\
\indent To obtain a best estimate for the separation distance of the CE, we took a high-resolution image of our trap using the CMOS camera. By normalizing the dimensions of the picture to the well known width of the electrodes, we estimated the gap between the CE to be about $2.391\pm0.085\,\text{mm}$, which gives a relative uncertainty of $3.54\%$. Using a portable microscope, one may be able to reduce the uncertainty of the separation distance to about $0.92\%$.\cite{ricci2019levitodynamics} We chose not to optimize this result any further since our $3.54\%$ suffices to demonstrate the particle classification to a desired level. This translates to a relative error of $5.59\%$ in the value of the electric field.\\

\end{document}